\documentclass[aps,showpacs,epsf,twocolumn,prb]{revtex4}
\usepackage{amssymb}
\usepackage{amsmath}
\usepackage{graphicx,psfrag}
\usepackage{braket}
\usepackage{float}
\usepackage{subfig}
\usepackage{tikz}
\usepackage{epstopdf}
\usepackage{pgfplots}
\usepackage[colorlinks=true, citecolor=blue, urlcolor = blue, linkcolor= red, bookmarks=true]{hyperref}

\begin{document}

\def \beq{\begin{equation}}
\def \eeq{\end{equation}}
\def \bse{\begin{subequations}}
\def \ese{\end{subequations}}
\def \bea{\begin{eqnarray}}
\def \eea{\end{eqnarray}}
\def \bem{\begin{displaymath}}
\def \eem{\end{displaymath}}
\def \bem{\begin{bmatrix}}
\def \eem{\end{bmatrix}}
\def \Ps{\hat{\Psi}(\boldsymbol{r})}
\def \Pds{\hat{\Psi}^{\dagger}(\boldsymbol{r})}
\def \i{{\int}d^2{\bf r}}
\def \bl{\bar{\boldsymbol{l}}}
\def \c{\hat{c}_{n,m}}
\def \cp{\hat{c}_{n',m'}}
\def \cd{\hat{c}_{n,m}^{\dagger}}
\def \cdp{\hat{c}_{n',m'}^{\dagger}}
\def \bb{\bibitem}
\def \nn{\nonumber}

\def \bs{\boldsymbol}
\def \hkx{\hat{k}_{x}}
\def \hky{\hat{k}_{y}}
\def \bq{\bar{q_{y}}}

\def \bc{\begin{center}}
\def \ec{\end{center}}

\title{\textbf{Hofstadter butterflies in magnetically modulated graphene bilayer: an algebraic approach }}
\author{ Manisha Arora$^{1}$, Rashi Sachdeva$^{2}$ and Sankalpa Ghosh$^{1}$}
\affiliation{$^{1}$ Department of Physics, Indian Institute of Technology Delhi, New Delhi-110016, India}
\affiliation{$^{2}$ Mathematical Physics and NanoLund, Lund University, Box 118, 22100 Lund, Sweden}

\begin{abstract}
It has been shown that Bernal stacked bilayer graphene (BLG) in a uniform magnetic field demonstrates integer quantum Hall effect with a zero Landau-level anomaly \cite{Geimbilayer}. In this article we consider such system in a two dimensional periodic magnetic modulation with square lattice symmetry.
It is shown algebraically that the resulting Hofstadter spectrum can be expressed in terms of the corresponding spectrum of monolayer graphene in  a similar magnetic modulation. In the weak-field limit, using the tight-binding model, we also derive the Harper-Hofstadter equation for such BLG system in a periodic magnetic modulation. We further demonstrate the topological quantisation of Hall conductivity in such system and point out that the quantised Hall plateaus are equally spaced for all quantum numbers for the quantised Hall conductivity.
\end{abstract}

\maketitle

\newpage
\section{Introduction}
A two-dimensional electron gas in a transverse magnetic field and a periodic potential, shows fractal energy spectrum of the shape of a butterfly, first shown in the seminal work by D. Hofstadter \cite{Hofstadter}, thereby acquiring its name 
as the Hofstadter butterfly. Within linear response theory, the Hall conductivity of  this system can be identified with a topological invariant called Chern number defined over the topological space of the Brillouin zone \cite{TKNN, Kohmoto} corresponding to the periodic potential. 
This explains the perfect quantisation of Hall conductivity leading to the integer quantum Hall effect \cite{Klitzing}.  These two remarkable facets of the same system, namely topologically quantised Hall conductance and the fractal like energy spectrum, 
both stem from the magnetic translation symmetry \cite{Zak, Fischbeck, Florek, Tapash}, that generalises the lattice translational symmetry of Bloch electrons in the presence of 
time-reversal symmetry breaking magnetic field. To explore the domain of application of such magnetic translation symmetry, in our recent work \cite{Manisha}  we studied the charge carriers in monolayer graphene (MLG) \cite{Novoselov1, Gusynin} that are massless Dirac fermions in a periodically modulated magnetic field. It was pointed out that the magnetic translation symmetry of a system in a uniform magnetic field and  a periodic potential is retained when the magnetic field is periodically modulated. We showed that in such a system the energy spectrum is again given by Hofstadter butterfly like fractal structure and the Hall conductivity is topologically quantised.

A natural yet non-trivial question can be asked by seeking the consequences of application of such two dimensional magnetic modulation to the quasiparticles of Bernal stacked bilayer graphene (BLG), implying that two hexagonal lattices of MLG with inequivalent sites $A,B$ and $\tilde{A},\tilde{B}$ are placed on top of each other such that $\tilde{A}$ site of the top layer is directly above the $B$ sites of the bottom layer and the two layers are coupled through interlayer hopping \cite{Bernal}.  Such Bernal stacked BLG can be thought of as an intermediate step between monolayer graphene and different types of Moir\'e superlattices \cite{andp} that are realised in  van der Waals heterostructures such as  graphene-hexagonal boron-nitride (hBN)  heterostructure \cite{Hofstadter1, Hofstadter2, Hofstadter3}
or twisted bilayer graphene \cite{TBLG}, where in the first case the Hofstadter butterfly structures have already been demonstrated experimentally. This itself  provides a sufficient motivation to study the effect of two-dimensional periodic magnetic modulation on the quasiparticles of BLG. 
However more intriguing aspect in the present context is that the Bernal stacked BLG shows a different chirality exhibiting a $2\pi$ Berry phase and a dominantly parabolic behaviour with a finite mass, as opposed to a $\pi$ Berry phase  and linear dispersion for the massless
chiral quasiparticles in MLG \cite{McCann}. Consequently in a transverse magnetic field, double degenerate Landau levels are formed incorporating two different orbital states with same energy \cite{Falkobilayer}
and giving rise to a new kind of integer quantum Hall effect in bilayer graphene with a zero level anomaly in the Landau level structure \cite{Geimbilayer}.
The different nature of chirality of the quasiparticles also manifests itself, not only when the graphene bilayer is exposed to a uniform magnetic field, but also when we consider the transmission of such massive chiral fermions through scalar potential barriers\cite{KSN3, kleinexp, Tan} or through vector potential barriers created by localised magnetic field dubbed as magnetic barriers \cite{Masir, levitov2, Neetu}.

In this work we start by showing that the low energy hamiltonian for the BLG in a periodic magnetic modulation commutes with the magnetic translation operator for a uniform field like in the corresponding problem for the MLG. As a result it is possible to obtain algebraically the Hofstadter like spectrum for the BLG in such periodically modulated magnetic field. The existence of two layers is reflected with the doubling of the number of Hofstadter butterflies in the spectrum when compared to the corresponding cases of MLG under similar approximation. In the present case, the separation between such doubled Hofstadter butterflies is controlled by the strength of the interlayer hopping. The Hall conductivity in this case also turns out to be topologically quantised. However 
unlike the experimental discovery of quantum Hall effect in bilayer graphene in uniform transverse magnetic field \cite{Geimbilayer}, that shows a topologically quantised Hall conductivity with somewhat similar expression but with a zero Landau level (LL) anomaly, here the quantised Hall plateaus are equally spaced. We explain various aspects of this feature.
Accordingly the rest of the paper is organised in the following way. After introducing the periodic magnetic modulation in section \ref{model}, in  section \ref{fourband} we write the effective four band low energy hamiltonian of BLG in the presence of such magnetic modulation and algebraically solve the problem using suitable commutator relations between different components of the 
canonical momentum operator.  This is followed by a calculation involving the use of tight binding approximation to write the lattice hamiltonian in such periodic modulation and to show that it again comes in the form of standard Harper-Hofstadter equation.
We plot the spectrum as a function of strength of the magnetic modulation. In the next section \ref{2band}, we work out algebraically the same energy spectrum, though this time, by projecting the hamiltonian to the lowest two bands. Thereafter it's comparative analysis  with the results from the four-band case is done. In section \ref{topo} using results obtained 
from the four-band model, within linear response theory we show that the Hall conductance is topologically quantised and there exists a $0$ value in the quantum number. We finally conclude our work.

\section{Bilayer graphene in the presence of two dimensional (2D) periodic magnetic modulation}
\subsection{The model} \label{model} 
In effective mass approximation, the single particle hamiltonian for Bernal stacked BLG in an arbitrary magnetic modulation pointed along the transverse direction ($z$-axis) can be written as \cite{Zarenia1} 
\beq H = 
   \begin{bmatrix}
     0 & \Pi & t & 0 \\
     \Pi^{\dagger} & 0 & 0 & 0 \\
     t & 0 & 0 & \Pi^{\dagger} \\
     0 & 0 & \Pi & 0  
   \end{bmatrix} \label{4band} 
\eeq 
giving the following eigenvalue  equation 
\beq
v_{F} H \Psi=E \Psi
\eeq 
where $\Psi=\begin{bmatrix} \psi_{a} & \psi_{b} & \psi_{c} & \psi_{d}   \end{bmatrix} ^{T}$. Here $t\approx 400~\text{meV} $ is the interlayer coupling, $v_{F}$ is the Fermi velocity, and 
\beq \Pi = \Pi_{x} - i \Pi_{y}, ~\Pi^{\dagger} = \Pi_{x} + i \Pi_{y}, \nonumber \eeq where the canonical momentum operator 
$\Pi_{x,y} =(p_{x,y}+\frac{eA_{x,y}}{c})$. The vector potential $\bs{A}(x,y)$ satisfies
\beq \bs{\nabla} \times \bs{A}(x,y) = \bs{B}(x,y) \hat{z}. \label{B} \eeq 
For a periodic magnetic field profile $\bs{B}(x,y)$ in the $x-y$ plane, 
\beq
\bs{B}=\sum_{\bs{G}} \bs{B}_{\bs{G}} e^ {i \bs{G}.\bs{r}} =\bs{B}_{u} + \bs{B}_{p} \label{Fourier}
\eeq
Here $\bs{G}$ is the reciprocal lattice vector of the periodic lattice, $\bs{B}_{u}$ is the uniform part of the magnetic field  given by the spatial average of the field given in Eq. $(\ref{B})$. The residual periodic magnetic field is defined as $\bs{B}_{p}= \sum_{\bs{G}\neq 0} \bs{B}_{\bs{G}} e^ {i \bs{G}.\bs{r}} $ that satisfies $
\bs{\nabla} \cdot \bs{B}_{p}=0$.  It can be easily checked that the net flux through the unit cell due to $\bs{B}_{p}$ is zero and hence comes entirely from $\bs{B}_{u}$. The results that are reported in this paper are generally true for any arbitrary periodic magnetic modulation in a two dimensional plane.
However for the sake of computation we choose a specific form of two dimensional periodic magnetic modulation. 
\begin{figure}[!htbp]
\begin{center}
\includegraphics[width=10cm,height=10cm,keepaspectratio]{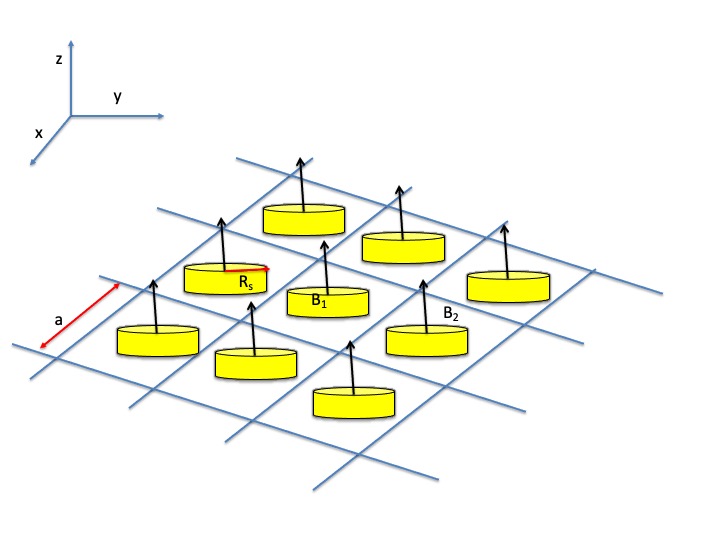}
\caption{{\it (Color online)}: Schematic plot of the magnetic field profile given in Eq. (\ref{B4}). $B_{1}$ and $B_{2}$ are the values of the field in the colored and uncolored regions in an arbitrary unit. The height of the coloured area is indicative of the fact that $B_{1} \neq B_{2}$.}
\label{Fig00}
\end{center}
\end{figure}

Our chosen two dimensional periodic magnetic field modulation is given by 
\beq \bs{B}= \Big[ B_{2}  + (B_{1} - B_{2}) \sum_{m,n} \Theta (R_{s} - r_{mn}) \Big] \bs{\hat{z}} \label{B4}\eeq 
Here  $R^{s}$ is radius of the circular region within each unit cell and $\bs{r}_{mn} = \bs{r} - \bs{R}_{mn}$
where $\bs{r}  =  x\bs{\hat{x}} + y\bs{\hat{y}}$. Apart from this, the lattice vector $\bs{R}_{mn}  = ma \bs{\hat{x}} + na\bs{\hat{y}}$
and 
\bea 
\Theta (R_{s} - r_{mn})
& = & 
\begin{cases}
1, ~~
\text{if}~\ r_{mn} \leq R_{s}, \\
0, 
 ~~\text{if}~\ r_{mn} > R_{s} \end{cases}   \eea 
where 
$m,n \in I (-N_{max}, +N_{max})$ and $a$ denotes the lattice constant. We assume that the lattice constant $a$ is much larger than the carbon-carbon (c-c) bond length in bilayer graphene which justifies the consideration of continuum model for bilayer graphene. A schematic presentation of the magnetic field profile is given in Fig. \ref{Fig00}. 
In general for any periodic magnetic field profile, because of the decomposition given in Eq. (\ref{Fourier}), the vector potential can be decomposed as
\beq \bs{A} = \bs{A}_{u} + \bs{A}_{p} \nonumber \eeq
with $A_{x}=A_{ux}+A_{px}, ~~A_{y}=A_{uy}+A_{py}$, where in a suitable gauge $\bs{A}_{p}$ can be written as a periodic function.
Accordingly, for the current magnetic field profile given in Eq. (\ref{B4}) we have chosen a vector potential that has two parts, namely 
 \bea
 \bs{A}_{u}(\bs{r}) & = & \frac{1}{2} \bs{B}_{u} \times \bs{r}, ~ \text{Symmetric Gauge}  \nn \\
 \bs{A}_{p} (\bs{r}) & = & \frac{1}{2} \sum_{m,n} \Big[(B_{p}^{mn} \bs{\hat{z}}) \times (r_{mn} \bs{\hat{r}}_{mn})\Big] \label{Ap} \eea
where
\begin{equation}
B_{p}^{mn}=
\begin{cases}
\Big[ (B_1 - B_2) - \frac{(B_1-B_2) \pi (R_{s}) ^{2}}{a ^2} \Big] \Theta(R_{s}-r_{mn}), \\
\text{if}\ r_{mn} \leq R_{s} \\\\
(B_{2}-B_{1}) \frac{\pi (R_{s})^2}{Na^2}\Theta(r_{mn} -R_{s}),~\text{otherwise}
\end{cases} \label{Bpmn}
\end{equation}
Because of the periodic structure of the magnetic modulation given in Eq.(\ref{B4}), the resulting quasiparticle spectrum of Bernal stacked BLG subjected to such periodic magnetic modulation will be fundamentally different from the corresponding four band dispersion of the BLG without such magnetic modulation \cite{McCann}.
The uniform field $\bs{B}_{u}$ in the expression (\ref{Fourier}) will lead to Landau levels  for such BLG \cite{Zarenia1}, whereas the additional $\bs{B}_{p}$ field which forms a lattice of the magnetic modulation will convert these Landau levels into Hofstadter bands \cite{Hofstadter}. It may be recalled that unlike in Moir\'e pattern in TBLG where a Hofstadter pattern arises \cite{MoireA, MoireB} due to the presence of the Moir\'e periodicity that results from the relative twist between the two layers each of which is individually treated in continuum, in the present problem the periodic structure is provided by the periodic modulation of the magnetic field itself given in Eq. (\ref{B4}).

\subsection{Energy spectrum using the four-band model in the presence of 2D periodic magnetic modulation} \label{fourband}

The eigenvalue equation for the hamiltonian  given in (\ref{4band})  leads to the following set of coupled equations 
 \bea v_{F} \Pi \psi_{b}  & = & E \psi_{a}-  v_{F} t\psi_{c} \label{1} \\
v_{F} \Pi^{\dagger} \psi_{a} & = & E \psi_{b} \label{2} \\
v_{F} \Pi^{\dagger} \psi_{d} & = & E\psi_{c}-  v_{F} t \psi_{a} \label{3} \\
v_{F} \Pi \psi_{c} & = & E \psi_{d} \label{4}
\eea 

\begin{figure}[!htbp]
\begin{center}
\includegraphics[width=14cm,height=11cm,keepaspectratio]{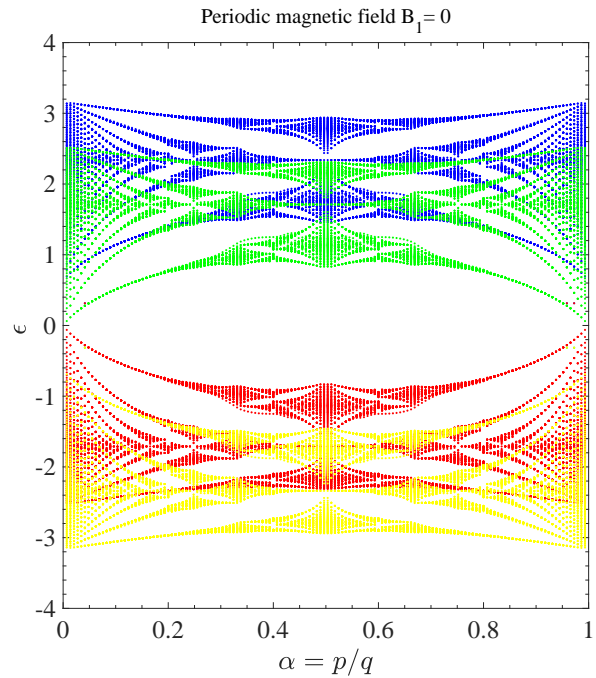}
\caption{{\it (Color online)}: Energy eigenvalues $\epsilon$ vs  flux $\alpha$ plot for $B_{1}=0$. $\epsilon=\frac{a E}{\hbar v_{F}}$. In this figure as well as in subsequent figures $\alpha = \frac{p}{q}$ represents the magnetic flux through the unit cell and $(p,q)$ are co-prime. For a fixed $B_{1}$,  
$B_{2}$ changes with change of $\alpha$ according to Eq. (\ref{alphab12}).}
\label{Fig10}
\end{center}
\end{figure}

We define $H_{G}= v_{F}^{2}\Pi \Pi^{\dagger}$. Since $H_{G}^{\dagger} = H_{G}$ coupling Eqs. (\ref{1}) - (\ref{2}), and Eqs. (\ref{3})-(\ref{4}) we get
\bea (H_{G} - E^{2}) \psi_{a}  & = & -v_{F} t \psi_{c} \label{5} \\
        (H_{G} -E^{2}) \psi_{c} & =& -v_{F}tE \psi_{a} \label{6} \eea 
The operator $H_{G}$ determines the energy spectrum for the monolayer graphene in the same periodic magnetic modulation. 
This tells us that $\psi_{a,c}$ will be obtained by solving the same eigenvalue equation upto an arbitrary phase factor like the MLG in a periodic magnetic modulation and $\psi_{b,d}$ can be obtained from Eqs. (\ref{2}) and (\ref{4}) subsequently for a given eigenvalue $E$.
This is one of the major findings of this work, and has consequences like the topological quantisation of the Hall conductivity in magnetically modulated BLG which is studied in details in subsequent sections. 

To get $\psi_{a,c}$ we combine Eqs. (\ref{5}) and (\ref{6}), giving 
\beq (H_G - E^{2})^{2} \psi_{a,c} = v_{F}^{2} t^{2} E^{2} \psi_{a,c}  \label{4bandeig} \eeq 
Square-rooting of the above equation directly gives us the the eigenvalue spectrum  $E$ of the BLG in the magnetic modulation defined in (\ref{B4}) as 
\beq
\epsilon=\frac{1}{2} \bigg{[}\pm \frac{a t}{\hbar} \pm \sqrt(\frac{a^2 t^2}{\hbar^2} + 4 \epsilon_{M})\bigg{]} \label{spectrumBLG} 
\eeq
where $\epsilon=\frac{a E}{\hbar v_{F}}$ and  $\epsilon_{M}=\frac{a^2 E_{M}^2}{\hbar^2 v_{F}^2}$ are dimensionless energies in BLG and MLG respectively with $E_{M}$ representing the energy of the MLG modulated by the same magnetic modulation.It can be readily checked that for $t=0$, the expression 
(\ref{spectrumBLG}) reduces to the corresponding spectrum of MLG in a way similar to the corresponding problem in uniform magnetic field case \cite{Zarenia1}.
The plots for the spectrum obtained using the above set of equations is shown in Fig. (\ref{Fig10})-  Fig. (\ref{Fig13}). 
In these plots the $x$-axis corresponds to $\alpha = \frac{p}{q}$ which is the total number of flux quanta through a unit cell and is given in terms of the co-prime $(p,q)$. $B_{1,2}$ defined in Eq. (\ref{B4}) is 
related to $\alpha$ through 
\beq \alpha \phi_{0} = B_{1} \pi R_{s}^{2} + B_{2}(a^{2} - \pi R_{s}^{2}). \label{alphab12} \eeq  
And the $y$-axis corresponds to the dimensionless energy eigenvalue. 

For a detailed analysis of the plots, we rewrite the LHS of Eq. (\ref{spectrumBLG}) as $\epsilon_{+-  ,  +-}$ where the first and second sign show the respective signs in the RHS of the  spectrum. 
This implies that, $\epsilon_{+  ,  +}$  and  $\epsilon_{- ,  -}$ correspond to the two outer butterflies (blue and yellow in Fig \ref{Fig10}) and $\epsilon_{+  ,  -} $ and  $\epsilon_{- ,  +}$ correspond to the two inner butterflies (green and red in Fig. \ref{Fig10})). 
For the case, when $\epsilon_M=0$, implying $\epsilon_{+,-}=\epsilon_{-,+}=0$, the two inner butterflies touch each other. This is the result obtained by setting interlayer coupling strength $t=400$ \text{meV} in Eq. (\ref{spectrumBLG}). 
\begin{figure}[!tbp]
  \centering
  \begin{minipage}[b]{0.40\textwidth}
    \includegraphics[width=\textwidth, height=\textwidth]{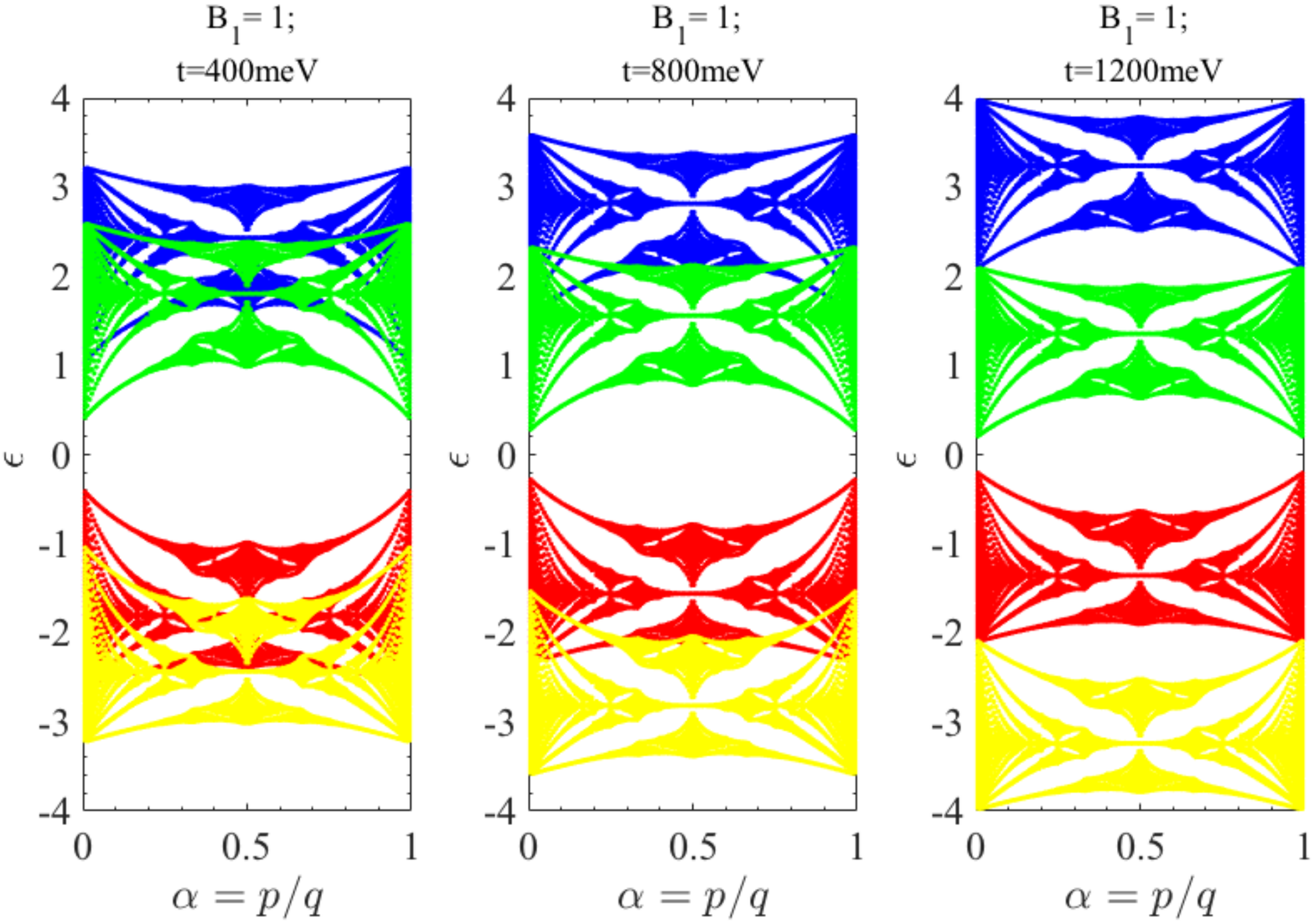}
    \caption{{\it (Color online)}: Energy spectrum $\epsilon$ defined in Eq. (\ref{spectrumBLG}) as a function of magnetic flux $\alpha$ for $B_{1}=1$ for three different values of interlayer coupling strengths $t$. The energy is plotted in the dimensionless unit $\epsilon=\frac{a E}{\hbar v_{F}}$.}
    \label{3t}
      \includegraphics[width=\textwidth, height= \textwidth]{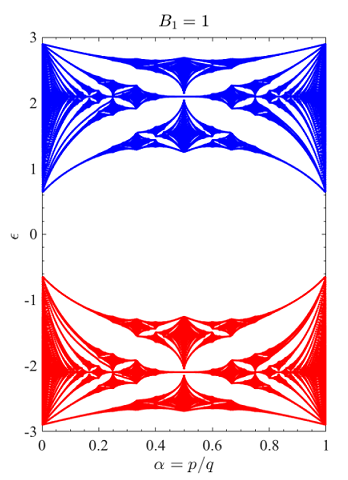}
    \caption{{\it(Color online)}: The corresponding energy spectrum  for monolayer graphene (MLG) as a function of magnetic flux $\alpha$ for the same value of $B_{1}$ as in Fig. \ref{3t}. Here $y$-axis corresponds to $\epsilon=\epsilon_{M}$ which is the dimensionless energy of MLG, and  is defined in the text after Eq. (\ref{spectrumBLG}).}
    \label{MLG}
  \end{minipage}
  \hfill
\end{figure}
\begin{figure} [!htbp]
\begin{center}
\includegraphics[width=11cm,height=8cm,keepaspectratio]{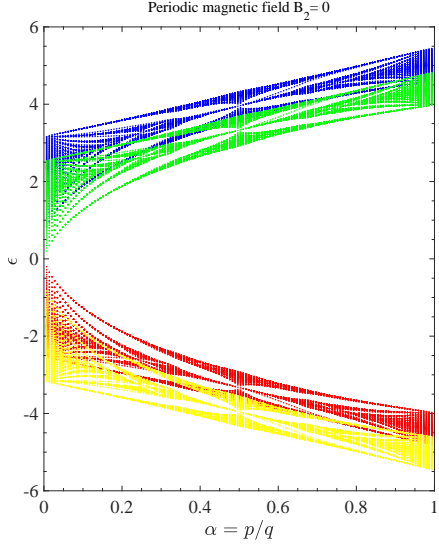}
\caption{{\it(Color online)}:Energy spectrum $\epsilon$ vs flux $\alpha$ plot for a fixed value of $B_{2}=0$, but varying $B_{1}$. As explained in the text, $\epsilon$ is the dimensionless energy. The relation between $B_{1}$ and $B_{2}$ is given in Eq. (\ref{alphab12}) that gives how $B_{1}$ varies with $\alpha$.}
\label{Fig12}
\end{center}
\end{figure}
\begin{figure} [!htbp]
\begin{center}
\includegraphics[width=11cm,height=8cm,keepaspectratio]{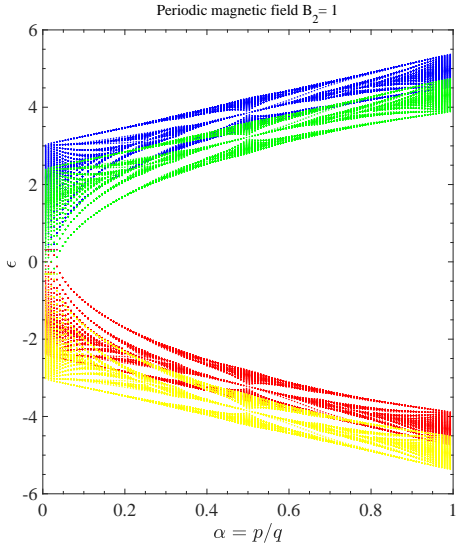}
\caption{{\it(Color online)}:$\epsilon$ vs. $\alpha$ plot for $B_{2}=1$. Notations are same as in Fig. \ref{Fig12}.}
\label{Fig13}
\end{center}
\end{figure}
As the value of $B_{1}$ is increased, the energy eigenvalue $\epsilon_{M}$ of the MLG in  a similar magnetic modulation increases (see Fig. (\ref{MLG})). 
This causes increase in the the gap between the outermost butterflies $\epsilon_{+,+}$  and  $\epsilon_{-,-}$. Similarly the gap between the inner butterflies $\epsilon_{+,-}$  and  $\epsilon_{-,+}$ also increases with increase in $B_{1}$. This can be seen in the plots as we move from Fig (\ref{Fig10}) $(B_{1}=0)$ as well as in the  leftmost figure of 
 Fig.  (\ref{3t}) $(B_{1}=1)$. 

We now study the effect of fixing the value of magnetic field strength $B_{1}$ and varying the interlayer coupling strength $t$. 
The effect is demonstrated in Fig. (\ref{3t}). 
 As $B_{1}$ is fixed, $\epsilon_{M}$ attains a fixed set of values.  Going from left to right  in Fig. (\ref{3t}), we can clearly observe that as t increases, the gap between the outer butterflies increases whereas the gap between the two inner butterflies decreases. 
 This can be understood from Eq. (\ref{spectrumBLG}), where as we increase t, the separation between $\epsilon_{+,+}$  and  $\epsilon_{-,-}$ increases whereas that between $\epsilon_{+,-}$  and 
 $\epsilon_{-,+}$ decreases. The spectrum is also plotted as a function of  fixed value of $B_{2}$, but varying values of $B_{1}$ in Fig. (\ref{Fig12}) and Fig. (\ref{Fig13}). 

Written in a slightly modified form, the eigenvalue equation for the BLG in a periodic magnetic modulation 
can be written in terms of the corresponding hamiltonian operator for the MLG as 
\beq H_{G} \psi_{a,c} = \big{[}(E^{\pm})^{2} - \frac{1}{4} v_{f}^{2} t^{2}\big{]} \psi_{a,c}  \label{MLGBLG} \eeq 
where $E^{\pm} = E \pm \frac{1}{2}v_{f} t$. 
It is known that $H_{G}$ commutes with the magnetic translation operator \cite{Zak} corresponding to the uniform field part of the magnetic modulation, namely $B_{u}$. 
Hence the eigenfunctions of the $H_{G}$, which are  also the magnetic Bloch functions,  can be expanded  
in terms of localised Wannier functions in the presence of
uniform transverse magnetic field $\bs{B}_{u}$\cite{Luttinger, Wannier}. 
\beq \psi_{\bs{k}}^{a,c} = \sum_{i} g(\bs{R}_{i}) \exp\big{(}-i \frac{e\bs{A}_{u} \cdot \bs{R}_{i}}{\hbar c}\big{)} w_{0}( \bs{r} - \bs{R}_{i}) \label{wf} \eeq

To simplify further calculation here we set the condition $|\frac{ea^{2}}{\hbar c} (B_{u}+B_{p})|<<1 $.
For $B_{p}=0$, this implies $ |\frac{ea^{2}}{\hbar c} (B_{u})| \ll 1$  which leads to $a \ll l_{B_{u}}$. Here $l_{B_{u}}$ is the magnetic length which is defined in terms of the uniform part of the magnetic modulation. 
Thus this condition corresponds to a weak and slowly varying magnetic field \cite{Luttinger, exact}. 
This is a reasonable approximation as long as the modulation part $B_{p}$ ( represented by either the r.m.s value, or the peak value) is smaller than $B_{u}$, and hence, can be treated as a perturbative effect.
 This type of condition is generally used in lattice gauge theory \cite{Governale}. Under these approximations we can write the eigenvalue equation ($\ref{MLGBLG})$ as a discrete Schr\"odinger equation which takes the form of
Hofstadter-Harper equation
\begin{widetext}
\bea
\epsilon g(m,n) &=&e^{\frac{i ea}{\hbar c} (A_{ux} + A_{px})}  g(m+1,n) + e^{-\frac{i ea}{\hbar c} (A_{ux} + A_{px})}  g(m-1,n) +  e^{\frac{i ea}{\hbar c} (A_{uy} + A_{py})} g(m,n+1) \nn\\
&+& e^{-\frac{i ea}{\hbar c} (A_{uy} + A_{py})} g(m,n-1) - [\frac{e a^2}{\hbar c} (B_{u} +B_{p}) + 4] g(m,n) \label{finalharper}
\eea
\end{widetext}
with \beq \epsilon=-\frac{[(E^{\pm})^{2} - \frac{1}{4} v_{f}^{2} t^{2}] a^2}{v_{F}^2 \hbar^2 } \label{EBLG} \eeq 
and $g(m,n)=g(\bs{R_{mn}})$. 
\begin{figure}[!tbp]
  \includegraphics[width=12cm,height=9cm,keepaspectratio]{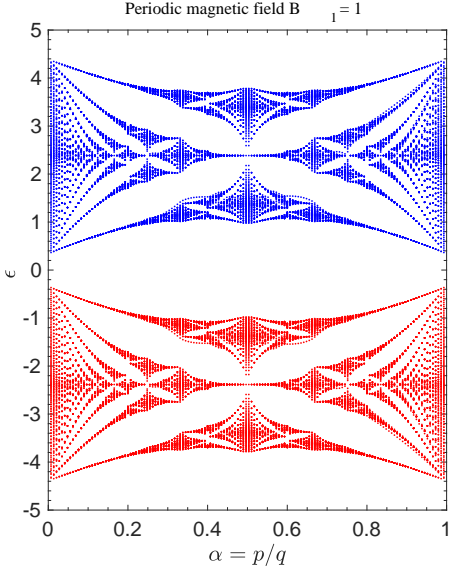}
    \caption{{\it(Color online)}:Bilayer graphene energy spectrum for $B_{1}=1$. The $y$-axis plots the dimensionless energy $\epsilon$ which is in the unit of $\frac{\hbar^{2}}{2ma^{2}}$. The $x$-axis plots the magnetic flux $\alpha=\frac{p}{q}$ like the previous figures.
    The expression of the energy is given in Eq. (\ref{EB}).}
    \label{Fig16}
\end{figure}
Once $\psi_{a,c}$ is calculated from Eqs. (\ref{wf}) and (\ref{finalharper}),  using Eq. (\ref{2}) we can calculate ( again under the weak magnetic field limit) 
\bea \psi_{\bs{k}}^{b} & = & \frac{v_{F}}{E} \bigg{[}- i \hbar (\frac{\psi_{\bs{k}} ^a(\bs{r} + a \bs{\hat{x}}) - \psi_{\bs{k}} ^a(\bs{r} - a \bs{\hat{x}})}{2a})\nn \\
&-& i \hbar (\frac{\psi_{\bs{k}} ^a(\bs{r} + a \bs{\hat{y}}) - \psi_{\bs{k}} ^a(\bs{r} - a \bs{\hat{y}})}{2a})\nn  \\
& + &  \frac{e}{c} (A_{u}^{x} + i A_{u}^{y} + A_{p}^{x} + i A_{p}^{y})\psi_{k}^{a}(\bs{r}) \bigg{]}  \eea 
In a similar manner $ \psi_{\bs{k}}^{d}$ also can be determined from  $\psi_{\bs{k}}^{c}$. 
Thus all the eigenfunctions can be determined. 
\subsection{Energy spectrum- Effective two-band model in 2D periodic magnetic modulation}\label{2band}
In the previous section we studied the BLG spectrum in a periodic magnetic modulation using the four-band model. In several BLG problems, an effective two-band model is used  \cite{Geimbilayer, Falkobilayer}
to analyse the behaviour of the charge carriers in BLG in an external field. In such two-band model, the low energy hamiltonian in effective-mass approximation is further projected on the lowest two energy bands \cite{McCann}. 
In this section we shall study the magnetically modulated BLG spectrum using this effective two-band model within the framework of  our algebraic approach. 

Using the form of the 
effective $2 \times 2$ hamiltonian at low energies for BLG charge carriers in the presence of 2d periodic magnetic modulation, we can write the eigenvalue equation as 
\begin{gather}
\frac{1}{2m} \left[ {\begin{array}{cc}
   0 & (\Pi^{\dagger})^2 \\
    (\Pi)^2 & 0 \\
 \end{array} } \right]  \begin{bmatrix} \psi_{1} \\ \psi_{2} \end{bmatrix} = E  \begin{bmatrix} \psi_{1} \\ \psi_{2} \end{bmatrix}
 \label{direq}
 \end{gather}
Using the commutator 
\beq
[\Pi^{\dagger}, \Pi]= \frac{2 \hbar e}{c} (B_{u} + B_{p}) \label{Comm}
\eeq
a straightforward decoupling yields (we retain the same notation for the hamiltonian and the energy eigenvalues) 
\beq
H \psi_{1}=E ^{2} \psi_{1} \label{eigval}
\eeq
with 
\begin{widetext} 
\bea
H&=& \frac{1}{4m^2} (\Pi^{\dagger})^2 (\Pi)^2 = \frac{1}{4m^2}(\Pi^{\dagger})\Big([\Pi^{\dagger}, \Pi] + \Pi\Pi^{\dagger}\Big)(\Pi) \nn \\
&=& \frac{1}{4m^2} \Big[ \frac{H_{G}^2}{v_{F}^4} +  \Big(\frac{2 \hbar e}{c} (B_{u}+ B_{p}) \Big) \frac{H_{G}}{v_{F}^2}  +  \frac{2 \hbar e}{c}\Big( \Pi^{\dagger} (B_{u} + B_{p}) \Big) \Pi \Big] \label{Hamm}
\eea
\end{widetext} 
Under the approximation that the magnetic field is  weak and slowly varying, we can ignore the last term appearing in the above hamiltonian which invokes the following argument. 
In the current problem $B_{u} + B_{p}=B(\vec{r})$ is spatially constant both inside and outside circular regions, whereas at the boundary of the two regions it is a step function and hence strictly speaking it's derivative will give discontinuity in the form of a delta function. But in reality there will be an intermediate region where the magnetic field will smoothly go from $B_{1}$ to $B_{2}$ over a finite region that is sufficiently large as compared to the bond-length of graphene, but sufficiently small compared to the relevant length scales that appear in the current problem. Therefore we can set 
\beq
\Big(p_{x}-ip_{y}\Big) \Big(B_{u} + B_{p}\Big) \approx 0
\eeq
throughout the considered region. 
Moreover, we also impose the condition $|\frac{e a^2}{\hbar c} B(\vec{r})| \ll1$ and 
$|\frac{e a A_{x,y} }{\hbar c}| \ll 1$. Given this 
\begin{widetext} 
\beq 
\frac{2 a^4 e}{\hbar^3 c} \Big( \Pi^{\dagger} (B_{u} + B_{p}) \Big) \Pi \\
=\frac{2 a^4 e}{\hbar^3 c} \Big[ \Big(p_{x}-ip_{y}\Big) \Big(B_{u} + B_{p}\Big) + \Big(\frac{e}{c} (A_{x} -i A_{y})\Big) \Big(B_{u} + B_{p}\Big)  \Big] \Pi  \approx =0  
\eeq
The hamiltonian $H$ can therefore be approximated as 
\beq H \approx \Big[ \frac{H_{G}^2}{v_{F}^4} +  \Big(\frac{2 \hbar e}{c} (B_{u}+ B_{p}) \Big) \frac{H_{G}}{v_{F}^2} \Big] = H' \eeq 
\end{widetext}
and the eigenvalue equation to be solved is 
\beq H' \psi_{1} = 4m^{2} E^{2} \psi_{1} \nonumber \eeq 
Since $[H', H_{G}]=0$, $\psi_{1}$ can be taken same as the eigenfunction of $H_{G}$. The other component of the spinorial wavefunction $\psi_{2}$
can be written as 
\beq \psi_{2} = \frac{1}{2mE} \hat{\Pi}^{2} \psi_{1} \eeq 
$E$ can now be determined as 
\beq E = \pm \frac{\hbar^{2}}{2m a^{2}} \sqrt { \epsilon_{M}^{2} + 4 \pi \frac{(B_{u} + B_{p})a^{2}}{\phi_{0}} \epsilon_{M}} \label{EB} \eeq 
$\epsilon_{M}$ is the dimensionless energy of the MLG in the same magnetic modulation.   
It is interesting to note that similar to the case of dispersion obtained from the effective two-band low energy description of BLG,
here also the energy scales in the unit of $\frac{\hbar^{2}}{2ma^{2}}$. We can therefore set the dimensionless energy here as $\epsilon = \frac{E}{\frac{\hbar^{2}}{2ma^{2}}}$. This  points out to the fact that the quasiparticles are massive, unlike the case of magnetically modulated MLG where the quasiparticles are massless. The energy spectrum is also plotted in the same unit in Fig. \ref{Fig16}. In this case we get two copies of the Hofstadter butterfly spectrum which is consistent with the two-band approximation. The analytical form of the energy spectrum in a magnetically modulated Bernal stacked BLG, derived respectively under four band and the two band approximation and given in Eq. (\ref{spectrumBLG}) and Eq. (\ref{EB}), forms one of the major result in this work.

\subsection{Approximation involved in obtaining the Hofstadter spectrum for magnetically modulated graphene layers}\label{approx}  
The spectrum presented in Figs. \ref{Fig10}, \ref{3t}, \ref{MLG}, \ref{Fig12}, \ref{Fig13}, \ref{Fig16} are all enumerated with the help of Eq. (\ref{finalharper}) which is a discrete eigenvalue equation. This is derived from the eigenvalue Eq. \ref{MLGBLG} 
by first expanding the solutions in terms of gauged Wannier functions defined in Eq. (\ref{wf})  and then using the nearest neighbour tight-binding approximation. As a result, for the MLG spectrum we obtained only two copies of the Hofstatdter butterflies both of which are bounded, and the BLG spectrum was obtained using the expression 
from the MLG spectrum by use of relation (\ref{spectrumBLG}) and Eq. (\ref{EB}). 
It may be recalled \cite{BookChap} that the eigenvalue equation (\ref{finalharper}) is similar in form to the celebrated Harper equation \cite{Harper} that was obtained in the original work of Hofstadter \cite{Hofstadter}, by considering the single lowest band of Bloch equation in the presence of a uniform magnetic field. 
In absence of the magnetic field, the band dispersion for such a single band in a square lattice periodic potential in the tight binding approximation is given by 
\beq E(k_{x}, k_{y}) = 2t ( \cos k_{x} a_{sl} + \cos k_{y} a_{sl} ) \label{square} \eeq 
where $k_{x}, k_{y}$ are Bloch momentums along $x,y$ directions respectively, $t$ is the energy scale set by hopping and $a_{sl}$ is the lattice constant of the square lattice. As can be seen from this expression,  
the width of such single band is finite, resulting in the finiteness of the energy spectrum given by the Hofstadter butterfly when the effect of a uniform magnetic field is added through Peierls' substitution.  
This situation may be contrasted carefully with the approach in the current problem where we start from the low-energy dispersion of Bernal-stacked bi-layer graphene or monolayer graphene in the $\bs{k} \cdot \bs{p}$ approximation that 
are respectively 
quadratic \cite{McCann} and linear \cite{Novoselov1} in quasi (Bloch)-momentum and hence unbounded.
When a  periodic  magnetic modulation given by Eq. \ref{Fourier} is added to such problem, 
it is expected that the resulting spectrum should show higher energy copies of the Hofstadter butterflies corresponding to the higher-energy part of the spectrum of the unperturbed problem. However, as can be seen, the spectrum presented in Figs. \ref{Fig10}, \ref{3t}, \ref{MLG}, \ref{Fig12}, \ref{Fig13}, \ref{Fig16} are all 
bounded.

The absence of these higher energy copies in our calculation can be explained by noting our 
restriction to the nearest neighbour tight-binding which is clearly visible in the form of Eq. (\ref{finalharper}). Here we first provide a brief justification for ignoring the higher order terms in the tight-binding approximation. Then we shall write some comments about how such restrictions can be relaxed and more accurate spectrum can be obtained.
As already mentioned that in the current problem we use the weak field limit $a \ll \ell_{B_{u}}$, where  $a$ accounts for the periodicity of the periodic magnetic modulation studied here. Even though the effective bilayer graphene hamiltonian is continuous and hence unbounded, the lattice present due to the periodic modulation justifies the discretisation 
and the nearest neighbour truncation of the resultant equation under this weak field limit. A single dispersion like Eq. (\ref{square}) for the current problem by considering only the $\bs{B}_{p}$ in Eq. \ref{Fourier} is also a non-trivial problem, because 
the lattice parameter $a$ here is not the carbon-carbon bond length, and is purely due to the magnetic modulation.
To extend the current calculation in the strong field limit (where the magnetic length is comparable or may be even lower than the lattice-spacing) one can follow the methods described in ref. \cite{Janecek} where the Scrh\"odinger equation in a periodic potential and magnetic field is numerically diagonalised exploiting the magnetic translational symmetry 
of the problem. Alternatively one can use the method adopted in ref. \cite{Gumbs} by first solving the corresponding problem in an uniform magnetic field \cite{Zarenia1}  and then applying the zero-flux periodic magnetic modulation represented by the fiield $B_{p}\hat{z}$ as a perturbation. 
Direct relaxation of our scheme can also be used, however a comparison between these various schemes is  not clear at this stage.
All these methods are numerically demanding and we hope 
to address some of them in future work.

\section{Topological quantization of Hall conductance for the modulated BLG- four band model}\label{topo} 
The Landau level spectrum in a transverse uniform magnetic field from the four band hamiltonian (\ref{4band}) is given by \cite{Zarenia1}
\beq E_{LL}^{\text{4band}} = s_{1} [ \frac{t^{2}}{2} +2N +1 +s_{2}\sqrt{\frac{t^{4}}{4} + (2N+1)t^{2} +1} ]^{\frac{1}{2}} \label{4bandLL} \eeq 
This expression may be compared with the expression (\ref{spectrumBLG}) that gives the corresponding spectrum under periodic magnetic modulation considered in the current work. 
Here $s_{1}=\pm{1}$ and $s_{2}=\pm{1}$ and $N$ is the Landau level index. The experimental results of quantised Hall conductivity in Bernal stacked bilayer graphene reported in the pioneering work \cite{Geimbilayer} 
was however theoretically analysed \cite{Falkobilayer, McCann} by the low energy two band hamiltonian and its Landau level spectrum for the AB (Bernal) stacked BLG in a transverse uniform magnetic field. 
The corresponding spectrum appears very different from the one given in (\ref{4bandLL}) and is given by 
\beq E = \pm \hbar \omega_{c} \sqrt{N(N-1)},~~N \ge 2 \label{2bandLL} \eeq 
where $N$ is the Landau level index and $E_{0} = E_{1}=0$. Particularly in the expression (\ref{2bandLL}) for zero energy eigenvalues thus there are two levels respectively for $N=0,1$. This causes the double degeneracy.
The valley structure and electronic spin degrees of freedom, each contributes a double degeneracy to the Landau level spectrum in (\ref{2bandLL}). Thus each level in bilayer graphene is four-fold degenerate except the zero energy level that shows an eight-fold degeneracy.
Accordingly  the quantized Hall conductivity in bilayer graphene \cite{Geimbilayer, Falkobilayer, Novoselov1} occurs at 
\beq
\sigma_{xy}=4N \frac{e^2}{h}
\eeq
\noindent
where $N \in Z$ but $N \neq 0$.  As has been pointed out \cite{McCann} that this quantization rule resembles that for 2DEG in a conventional semiconductor with degeneracy $4$ coming from the spin and valley degrees of freedom. But, with the exception that here $N=0$ plateau is missing and there is a step of height $\frac{8e^{2}}{\hbar}$  across the zero-electron density (the Hall conductivity being plotted at a fixed magnetic field). Additionally a crucial assumption in these works is that the degeneracy of the Landau level is preserved and any splitting of the Landau level ( such as Zeeman splitting) is negligible as compared to the level broadening ( due to disorder etc.) and temperature effect in the experiment.

A direct comparison between these mentioned results and that of ours, namely the fate of quantised Hall conductivity in the presence of periodic magnetic modulation is somewhat out of scope of the current work. This is due to the fact that the model hamiltonian (\ref{4band}) is based on a single valley and single spin. 
We assumed that the magnetic modulation given in Eq. (\ref{B4}) is sufficiently strong to freeze the spin-degrees of freedom and there is no term that allows scattering between the two valleys. Given this form if we extend the treatment of the TKNN model \cite{TKNN, Kohmoto} to the current problem described by the hamiltonian 
(\ref{4band}) we can demonstrate the topological quantisation of the Hall conductivity for the same without any anomalous behaviour at the zero density as expected from the spectrum given in (\ref{spectrumBLG}). 
Accordingly in this section, we shall derive the topological quantization of Hall conductivity for the BLG under two dimensional periodic magnetic field, using the four band model. Extension of this treatment  using an effective two-band model described in section \ref{2band} and including spin and valley degeneracy is not considered in the current work.

For the magnetic modulation defined in Eq. (\ref{B4}),  the eigenfunctions of the hamiltonian (\ref{4band}) are simultaneous eigenfunctions for the magnetic translation group 
defined for the uniform magnetic field part $B_{u}$ of the magnetic modulation, and hence can be written as 
\begin{gather}
 \begin{bmatrix} \psi^{a1}_{k_{x}, k_{y}} \\ \psi^{b2} _{k_{x}, k_{y}} \\ \psi^{a2}_{k_{x}, k_{y}} \\ \psi^{b1}_{k_{x}, k_{y}} \end{bmatrix}
 =
 e^{i \bs{k}.\bs{r}}
  \begin{bmatrix}
   u^{a1}_{k_{x}, k_{y}} \\
   u^{b2}_{k_{x}, k_{y}}  \\
   u^{a2}_{k_{x}, k_{y}} \\
   u^{b1}_{k_{x}, k_{y}}
   \end{bmatrix} \label{psiu4}
\end{gather}
Since $B_{u} \neq 0$, the commutation relation between the magnetic translation operator ensures that each magnetic unit cell contains zero's of such wave-function and thus \cite{Kohmoto} 
\beq u^{a,b}_{k_{x}, k_{y}}(x,y)   = |u^{a,b}_{k_{x}, k_{y}} |e^{i \theta^{a,b}_{k_{x}, k_{y}}}  \eeq 
where the functions $u^{a,b}_{k_{x}, k_{y}}(x,y)$ can be found out by the solving the Schr\"odinger equation and the full four-component wave-function is normalised to $1$.
\beq H(k_{x}, k_{y}) u^{n_{b}}_{k_{x}, k_{y}} = E^{n_{b}} u^{n_{b}}_{k_{x}, k_{y}}  \eeq 
with $n_{b}$ being the band-index. For the current problem we solve for a single band and hence the band index is omitted. Rest of the derivation that will lead to the expression of Hall conductivity is very similar to the
one carried out  for the non-relativistic two dimensional gas in the TKNN work \cite{Kohmoto} or the one done for MLG in such magnetic modulation \cite{Manisha}, but with an important difference coming from the 4-component 
pseudo-spinorial nature of the magnetic Bloch function. It may be noted that this four component nature of the magnetic Bloch function is due to the existence of the two layers (each having two sublattices conventionally known as $A$ and $B$) in the bilayer graphene \cite{McCann} and not related with spin or valley degeneracy.
This directly follows from the hamiltonian (\ref{4band}) which is written for single valley and single spin.
 We therefore provide here some essential steps of the derivation for the Hall conductivity involving this four component pseudo-spinor for the magnetically modulated BLG 

To this purpose we first define
$u'_{k_{x}, k_{y}}=\begin{bmatrix} u^{a1}_{k_{x}, k_{y}} & u^{b2}_{k_{x}, k_{y}} & u^{a2}_{k_{x}, k_{y}} & u^{b1}_{k_{x}, k_{y}} \end{bmatrix}^{T}$.
\noindent
Within the linear response regime, the Hall conductance given by Kubo formula for a filled single band can be written as 
\beq
\sigma_{xy}=\frac{e^2}{h} \frac{1}{2 \pi i} \int d^2 k [\bs{\nabla_{k}} \times \bs{\hat{A}}(k_{x}, k_{y}) ] _{3} \label{5014}
\eeq 
where the integration is over the magnetic Brillouin zone (MBZ) and 
\beq
\bs{\hat{A}} (k_{x},k_{y})=\int d^2 r u'^*_{k_{x}, k_{y}} \bs{\nabla_{k}} u'_{k_{x}, k_{y}} \label{bconn}
\eeq
Using Stoke's law, Eq. $(\ref{5014})$ yields 
\beq
\sigma_{xy}=\frac{e^2}{h} \frac{1}{2 \pi i} \int_{\partial MBZ} \bs{dk}.\bs{\hat{A}} (k_{x}, k_{y}) \label{sig34} \eeq 
To understand the non trivial topology of $\bs{\hat{A}} (k_{x},k_{y})$, we consider the gauge transformation
\begin{gather}
 \begin{bmatrix}
         u^{a1a}_{k_{x}, k_{y}} \\
           u^{b2b}_{k_{x}, k_{y}} \\
            u^{a2a}_{k_{x}, k_{y}} \\
           u^{b1b}_{k_{x}, k_{y}}
         \end{bmatrix}
 =
 e^  {i f(k_{x}, k_{y})}
\begin{bmatrix}
         u^{a1}_{k_{x}, k_{y}} \\
           u^{b2}_{k_{x}, k_{y}} \\
            u^{a2}_{k_{x}, k_{y}} \\
           u^{b1}_{k_{x}, k_{y}}
         \end{bmatrix}  \label{transf4}
\end{gather}         
         where $f(k_{x}, k_{y})$ is some arbitrary smooth function of only $k_{x}$ and $k_{y}$. We divide the MBZ into two regions, one contains the zero of the wavefunction and other does not. For region I, we can have
\[
\begin{bmatrix}
         u^{a11}_{k_{x}, k_{y}} \\
           u^{b21}_{k_{x}, k_{y}} \\
           u^{a21}_{k_{x}, k_{y}} \\
           u^{b11}_{k_{x}, k_{y}}
         \end{bmatrix}
=e^  {i g(k_{x}, k_{y})}
\begin{bmatrix}
         u^{a1}_{k_{x}, k_{y}} \\
           u^{b2}_{k_{x}, k_{y}} \\
            u^{a2}_{k_{x}, k_{y}} \\
           u^{b1}_{k_{x}, k_{y}}
         \end{bmatrix}         
         \] \label{u14}
and for region II
\[
\begin{bmatrix}
         u^{a12}_{k_{x}, k_{y}} \\
           u^{b22}_{k_{x}, k_{y}} \\
           u^{a22}_{k_{x}, k_{y}} \\
           u^{b12}_{k_{x}, k_{y}}
         \end{bmatrix}
=e^  {i h(k_{x}, k_{y})}
\begin{bmatrix}
         u^{a1}_{k_{x}, k_{y}} \\
           u^{b2}_{k_{x}, k_{y}} \\
           u^{a2}_{k_{x}, k_{y}} \\
           u^{b1}_{k_{x}, k_{y}}
         \end{bmatrix}         
         \] \label{u24}
Hence Eq. $(\ref{5014})$ becomes 
\bea \sigma_{xy}&=&\frac{e^2}{h} \frac{1}{2 \pi i} \Big[ \int_{H_{1}} d^2 k [\bs{\nabla} \times \bs{\hat{A}}^1 (k_{x}, k_{y}) ] _{3} \nn\\
&+&  \int_{H_{2}} d^2 k [\bs{\nabla} \times \bs{\hat{A}}^2 (k_{x}, k_{y}) ] _{3} \Big] \label{5024}
\eea
where the vectors $\bs{\hat{A}}^1 (k_{x},k_{y})$ and $\bs{\hat{A}}^2 (k_{x},k_{y})$ are defined for regions I and II respectively, and are given as 
\begin{widetext}
\bea
\bs{\hat{A}}^1 (k_{x},k_{y})&=& \int d^2 r [\bs{\hat{k_{x}}} (i \frac{\partial g}{\partial k_{x}}  u^{a1*}_{k_{x}, k_{y}} u^{a1}_{k_{x}, k_{y}} + i \frac{\partial g}{\partial k_{x}}  u^{b2*}_{k_{x}, k_{y}} u^{b2}_{k_{x}, k_{y}} + u^{a1*}_{k_{x}, k_{y}} \frac{\partial}{\partial k_{x}} u^{a1}_{k_{x}, k_{y}}  \nn\\
&+& u^{b2*}_{k_{x}, k_{y}} \frac{\partial}{\partial k_{x}} u^{b2}_{k_{x}, k_{y}} + i \frac{\partial g}{\partial k_{x}}  u^{a2*}_{k_{x}, k_{y}} u^{a2}_{k_{x}, k_{y}} + i \frac{\partial g}{\partial k_{x}}  u^{b1*}_{k_{x}, k_{y}} u^{b1}_{k_{x}, k_{y}} + u^{a2*}_{k_{x}, k_{y}} \frac{\partial}{\partial k_{x}} u^{a2}_{k_{x}, k_{y}} \nn\\
&+& u^{b1*}_{k_{x}, k_{y}} \frac{\partial}{\partial k_{x}} u^{b1}_{k_{x}, k_{y}} ) + \bs{\hat{k_{y}}} ( i \frac{\partial g}{\partial k_{y}}  u^{a1*}_{k_{x}, k_{y}} u^{a1}_{k_{x}, k_{y}} + i \frac{\partial g}{\partial k_{y}}  u^{b2*}_{k_{x}, k_{y}} u^{b2}_{k_{x}, k_{y}} \nn\\
&+& u^{a1*}_{k_{x}, k_{y}} \frac{\partial}{\partial k_{y}} u^{a1}_{k_{x}, k_{y}}  + u^{b2*}_{k_{x}, k_{y}} \frac{\partial}{\partial k_{y}} u^{b2}_{k_{x}, k_{y}}+ i \frac{\partial g}{\partial k_{y}}  u^{a2*}_{k_{x}, k_{y}} u^{a2}_{k_{x}, k_{y}} + i \frac{\partial g}{\partial k_{y}}  u^{b1*}_{k_{x}, k_{y}} u^{b1}_{k_{x}, k_{y}} \nn\\
&+& u^{a2*}_{k_{x}, k_{y}} \frac{\partial}{\partial k_{y}} u^{a2}_{k_{x}, k_{y}}  + u^{b1*}_{k_{x}, k_{y}} \frac{\partial}{\partial k_{y}} u^{b1}_{k_{x}, k_{y}} )] \label{A14}
 \eea
and
\bea
\bs{\hat{A}}^2 (k_{x},k_{y})&=& \int d^2 r [\bs{\hat{k_{x}}} (i \frac{\partial h}{\partial k_{x}}  u^{a1*}_{k_{x}, k_{y}} u^{a1}_{k_{x}, k_{y}} + i \frac{\partial h}{\partial k_{x}}  u^{b2*}_{k_{x}, k_{y}} u^{b2}_{k_{x}, k_{y}} + u^{a1*}_{k_{x}, k_{y}} \frac{\partial}{\partial k_{x}} u^{a1}_{k_{x}, k_{y}} \nn\\
&+& u^{b2*}_{k_{x}, k_{y}} \frac{\partial}{\partial k_{x}} u^{b2}_{k_{x}, k_{y}} + i \frac{\partial h}{\partial k_{x}}  u^{a2*}_{k_{x}, k_{y}} u^{a2}_{k_{x}, k_{y}} + i \frac{\partial h}{\partial k_{x}}  u^{b1*}_{k_{x}, k_{y}} u^{b1}_{k_{x}, k_{y}} + u^{a2*}_{k_{x}, k_{y}} \frac{\partial}{\partial k_{x}} u^{a2}_{k_{x}, k_{y}} \nn\\
&+& u^{b1*}_{k_{x}, k_{y}} \frac{\partial}{\partial k_{x}} u^{b1}_{k_{x}, k_{y}})+\bs{\hat{k_{y}}} ( i \frac{\partial h}{\partial k_{y}}  u^{a1*}_{k_{x}, k_{y}} u^{a1}_{k_{x}, k_{y}} + i \frac{\partial h}{\partial k_{y}}  u^{b2*}_{k_{x}, k_{y}} u^{b2}_{k_{x}, k_{y}} \nn\\
&+& u^{a1*}_{k_{x}, k_{y}} \frac{\partial}{\partial k_{y}} u^{a1}_{k_{x}, k_{y}}  + u^{b2*}_{k_{x}, k_{y}} \frac{\partial}{\partial k_{y}} u^{b2}_{k_{x}, k_{y}} + i \frac{\partial h}{\partial k_{y}}  u^{a2*}_{k_{x}, k_{y}} u^{a2}_{k_{x}, k_{y}} + i \frac{\partial h}{\partial k_{y}}  u^{b1*}_{k_{x}, k_{y}} u^{b1}_{k_{x}, k_{y}} \nn\\
&+& u^{a2*}_{k_{x}, k_{y}} \frac{\partial}{\partial k_{y}} u^{a2}_{k_{x}, k_{y}}  + u^{b1*}_{k_{x}, k_{y}} \frac{\partial}{\partial k_{y}} u^{b1}_{k_{x}, k_{y}} )] \label{A24}
\eea 
\end{widetext}
Since the full four-component spinorial wavefunction in Eq. (\ref{psiu4}) is normalised, a straightforward algebra now gives 
\beq
\bs{\hat{A}}^1 (k_{x},k_{y})-\bs{\hat{A}}^2 (k_{x},k_{y})= i \bs{\nabla_{k}} t(\bs{k}) \label{5044}
\eeq 
where $t(\bs{k})=g(\bs{k})-h(\bs{k})$.

We now write, using Stokes theorem and the fact that the two regions have opposite directions for circulation, the Eq. $(\ref{5024})$ as 
\beq
\sigma_{xy}=\frac{e^2}{h} \frac{1}{2 \pi i} \int_{C} \bs{dk}. ( \bs{\hat{A}}^1 (k_{x}, k_{y})- \bs{\hat{A}}^2 (k_{x}, k_{y})) \nn
\eeq
which on using Eq. $(\ref{5044})$ leads to 
\beq
\sigma_{xy}=\frac{e^2}{h} \frac{1}{2 \pi } \int_{C} \bs{dk}. \bs{\nabla_{k}} t(\bs{k}) \label{sigma42}
\eeq 
where $C$ denotes closed boundary between the two regions.
At each point on the closed loop C, the wavefunction has to be single valued therefore after traversing the complete loop C the two wavefunctions still should have the same phase relationship. This is possible only if
\beq
\begin{bmatrix}
         u^{a12}_{k_{x}, k_{y}} \\
           u^{b22}_{k_{x}, k_{y}} \\
           u^{a22}_{k_{x}, k_{y}} \\
           u^{b12}_{k_{x}, k_{y}}
         \end{bmatrix}
=e^  {i (t(k_{x}, k_{y})+ 2 \pi \sigma_{H})}. 
\begin{bmatrix}
         u^{a11}_{k_{x}, k_{y}} \\
           u^{b21}_{k_{x}, k_{y}} \\
           u^{a21}_{k_{x}, k_{y}} \\
           u^{b11}_{k_{x}, k_{y}}
         \end{bmatrix}
          \nn    
            \eeq
Thus expression $(\ref{sigma42})$ finally comes out to be 
\beq
\sigma_{xy}=\frac{e^2}{h} \sigma_{H} \label{TKNNBLG} 
\eeq  
\begin{figure} [!htbp]
\includegraphics[width=10cm,height=10cm]{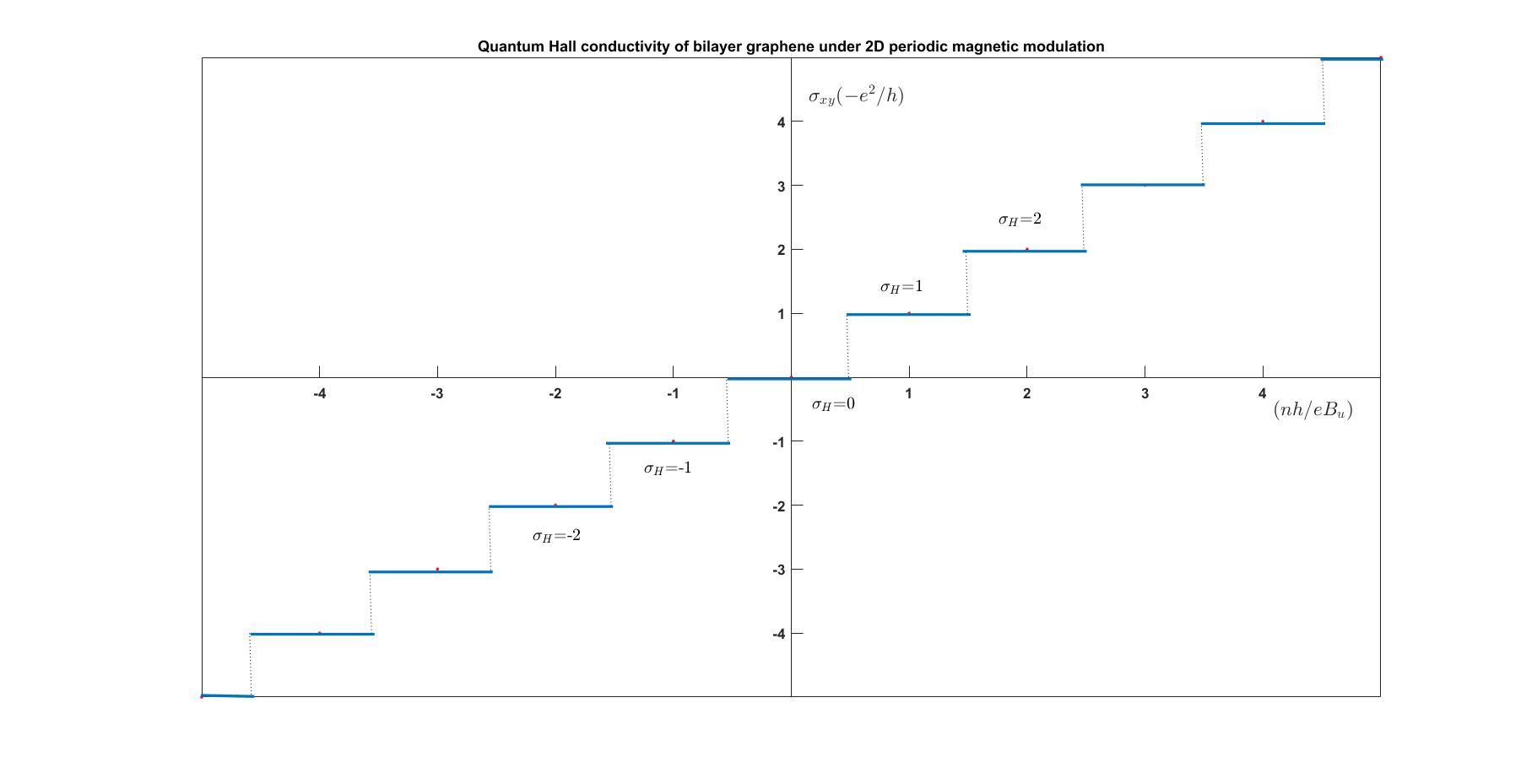}
 \caption{Quantum Hall plateaus in bilayer graphene in periodic magnetic modulation at zero temperature: The y-axis plots the quantised Hall resistivity and the $x$-axis plots the magnetic field strength $B_{u}$ at a fixed electron density $n$. All other scaling factors are shown in the axis label and have their usual meaning. }
     \label{sigmaH}
\end{figure}

where $\sigma_{H} \in Z$. Some remarks on this finding of our work are in order. As explained earlier, this four component nature of the wave function stems from 
the existence of two layers each having two sub-lattices. We have not considered the effect of the valley and spin in our model. Unlike in the case of the Landau level spectrum given by the two band hamiltonian (\ref{2bandLL}) in the presence of a uniform transverse magnetic field considered 
in  \cite{Falkobilayer, McCann2}, the relevant energy spectrum for the four band model given in (\ref{spectrumBLG})
does not distinguish between the $0,1$ LL index with the rest. As a result the quantised Hall plateaus in our case are always equally spaced. 
In this connection it may be pointed out that the lifting of the degeneracy of $0,1$ LL index was earlier predicted  with the application 
of an out of plane electric field to the bilayer graphene in addition to the transverse magnetic field and was observed experimentally \cite{Geim3}. The current situation is  different from that.
Thus in a way quite similar to the case of two-dimensional electron system 
in a uniform magnetic field and a periodic lattice potential \cite{TKNN, Kohmoto}, if the Fermi energy lies in one of the gaps of the spectrum, the Hall conductivity is quantised in terms of an integer. 
Our treatment of evaluating quantised Hall conductivity is also different from the method adopted in ref. \cite{Falkobilayer}, where they have used the two-band model to evaluate the the quantised Hall effect. However we expect that even in that calculation, the periodic magnetic modulation should lift the zero level anomaly. This expectation is based  on the expression 
of the energy spectrum given in (\ref{EB}) within the two-band model in the presence of the periodic modulation that does not show any $N=0$ degeneracy that appears when the field is uniform. 
We have not considered this problem any further here.
Such topological quantisation is naturally robust against effect of disorder or any other perturbation as long as it does not close the gap. As a notional demonstration we have plotted such quantised Hall conductivity plateaus given in Eq. (\ref{TKNNBLG}) at absolute zero temperature in Fig. (\ref{sigmaH}). Please note the plotted quantities are not directly calculated from Kubo formula, within linear response regime, but the plot of  $\sigma_{H}$ given in 
Eq. (\ref{TKNNBLG}) at the corresponding values of $B_{u}$. The construction of the plateaus are done just by connecting the intermediate regions with lines parallel to $x$-axis. Under ideal situation this represents the zero temperature quantised Hall conductivity plot and hence the phrase "notional demonstration" . 

To summarise, we demonstrated rigorously that the quantised Hall conductance can survive in Bernal stacked bilayer graphene when the magnetic field is periodically modulated. We pointed out that this is a result of the preservation of magnetic translational symmetry in the resulting system, and provides some algebraic results for the spectrum. We have pointed out that in the framework of considered theoretical model in this work, the resulting Hall conductivity in our system is quantised in the unit of $\frac{e^2}{h}$ for all quantum numbers for the quantised Hall conductivity, unlike the anomalous behaviour of the quantised Hall conductivity in bilayer graphene \cite{Geimbilayer} in uniform magnetic field due to a degeneracy between $0,1$ LL index.  
The type of magnetic modulation that has been considered in this work can be created in a number of ways.  Such techniques span from
the use of nano-lithography \cite{Bader} to more recent days structure such as two dimensional magnetic van der Waals materials \cite{Vanm} or by arranging ad-atoms of a ferromagnetic material in periodic arrays on a suitable surface \cite{adatom}, to name a few. 
The presence of periodic pseudo-magnetic field (superlattice) was experimentally confirmed in buckled monolayer graphene \cite{Mao}. The order of lattice constant in such superlattices is of the order of $40$ nm and the field strength is $\sim 100~\text{Tesla}$.
Several other recent experiments also realised graphene in periodic magnetic modulation by subjecting  graphene to strain induced superlattice \cite{strain}, or by exposing epitaxial graphene to an abrikosov vortex lattice created on a type II superconductor \cite{sc}. In the work \cite{strain}, it was demonstrated that periodic modulation of strain applied to graphene modulates the $C-C$ bond and gives rise to intense pseudo-magnetic field superlattice 
 of lattice constant $1-10$ nm and field strength $\sim 100~\text{Tesla}$. Where in some such cases, the weak field limit that we adopt in this work is approximately respected, in other cases the strength of the modulation is stronger. 
As discussied in sec. \ref{approx}, the nearest neighbour tight-binding approximation that is used in the current work need to be relaxed in the stronger field cases, which will show higher energy copies of the Hofstatder butterflies. 
We can therefore hope that our work may stimulate new experimental works and further theoretical analysis in future in similar systems.

\newpage
\def \bb{\bibitem}
\def \bb{\bibitem}

\end{document}